\documentclass[12pt]{article}
\usepackage{amsmath,amsthm,amsfonts,graphicx,epsfig}
\usepackage[usenames]{color}
\newcommand{\ket}[1]{\left| #1 \right\rangle}
\newcommand{\bra}[1]{\left\langle  #1 \right|}

\newcommand{\tr}{\mathop{\mathrm{tr}}}

\newcommand{\e}{\varepsilon}
\newcommand{\dbar}{\kern-.1em{\raise.8ex\hbox{ -}}\kern-.6em{d}}
\newcommand{\comment}[1]{}
  
\def\half{\mbox{$\frac 1 2$}}

\def\lin{{\cal L}}
\def\Ker{{\mathrm {Ker}}}
\def\Range{{\mathrm {Range}}}

\newtheorem{thm}{Theorem}%[section]

\def \be{\begin{equation}}
\def \ee{\end{equation}}
\def \bea{\begin{eqnarray}}
\def \eea{\end{eqnarray}}

\usepackage[usenames]{color}
\author{J.E. Avron, M. Fraas,
\\
\small{Department of Physics, Technion, 32000 Haifa, Israel}
\\ G.M. Graf,
P. Grech
\\
\small{Theoretische Physik, ETH Zurich, 8093 Zurich, Switzerland} }
\begin{document}
\title{Landau-Zener tunneling  for dephasing Lindblad evolutions }
\date{\today}%
\maketitle

\begin{abstract}
We consider a family of time dependent dephasing Lindblad
generators which model the monitoring of the instantaneous
Hamiltonian of a system by a Markovian bath. In this family the
time dependence of the dephasing operators is (essentially)
governed by the instantaneous Hamiltonian. The evolution in the
adiabatic limit admits a geometric interpretation and can be
solved by quadrature. As an application we derive an analog of the
Landau-Zener tunneling formula for this family.
\end{abstract} \maketitle

Lindblad generators describe the quantum evolutions of finite
systems coupled to a memoryless bath \cite{Davies}. They give a
useful phenomenological description of thermalization and
decoherence \cite{zoller,preskill}. We consider a family of
time-dependent dephasing Lindblad generators, first introduced in
\cite{ aks},  which models the
continuous monitoring of the instantaneous energy. This is the
case, for example, in the Zeno effect \cite{zeno}.  The family  can
be defined for arbitrary dephasing rate, however, its physical
interpretation in the strong dephasing regime requires some care
\cite{afgg}.

The family of dephasing Lindbladians that we consider is defined in
such a way that any instantaneous stationary states of the
Hamiltonian are also instantaneous stationary states for the
Lindbladian. This makes the family special and non-generic.
(Generic Lindbladians have a unique stationary state). In the
adiabatic limit, the evolution generated by this family can be
solved by quadrature and admits a geometric interpretation. The
qualitative features of the dynamics differ from the corresponding
dynamics of generic Lindbladians \cite{joye} reflecting the special
character of the family.

In 1932 Landau \cite{Landau} and independently Zener \cite{Zener}
and Majorana \cite{Majorana}
found an explicit formula for the tunneling in a {\em generic}
near-crossing undergoing unitary, adiabatic evolution. Here we
describe the corresponding result for the non-unitary case
associated with dephasing Lindbladian. The solution  appears to be
the simplest generalization of the Landau Zener problem which is
still exactly soluble.

The influence of dissipation and noise on the tunneling of a two
level system has been extensively studied in the physics
literature \cite{leggett,zenner-dissipation,shimshoni,noise}.
The results presented here differ in one or both of the following aspects: 
First, in the framework: We
\emph{assume} a Lindbladian, and do not attempt to derive an effective 
dynamics from
a (unitary) model of a bath or a (unitary) model of stochastic
noise source. Second: The Lindbladians are, as stated, of the dephasing type.
Our results contribute to the mathematical physics
of Lindblad operators. This approach has the virtue that the
adiabatic dynamics can be solved by quadrature and does not rely
on the assumption of weak dephasing which one normally needs to
make when modelling decoherence with a unitary bath or noise.

In the limit of weak dephasing, the tunneling formula we derive
can be compared with results of \cite{shimshoni} for Zener
tunneling due to a dephasing noise. The two formulas have the same
functional form up to an overall constant which is left
undetermined in \cite{shimshoni}. 

Since tunneling is dominated by the near crossing dynamics, the
universal aspects of near crossing in a two level system are
captured by a Hamiltonian that depends {\em linearly} on time. By
an appropriate choice of basis and of the zero of energy the
relevant dynamics is governed by the Hamiltonian
    \be\label{eq:H}
    H(s,g_0)=\half\,\left(%
\begin{array}{cc}
  s & g_0 \\
  g_0 & -s \\
\end{array}%
\right),\quad (s=\e t)
    \ee
where $\e>0$ is the adiabatic parameter and $g_0>0$ is the
minimal gap.
The tunneling probability $T$ is the probability of a state,
which originates asymptotically on one eigenvalue branch, to end up in the
other at late times.
The formula Landau and Zener found
\cite{history} for this Hamiltonian is:
     \be\label{eq:LZT}
    T= e^{-\pi g_0^2/2\hbar \e }.
    \ee
The singularity of the limit $\hbar \e\to 0$ reflects the
singularity of the adiabatic and semiclassical limits, and their
coincidence in this case.

The universal aspects  of tunneling for near crossing in an {\em open} system described by a dephasing Lindbladian is described as follows. 
The adiabatic evolution of the density matrix
$\rho$
is governed by
    \be\label{eq:L}
    \hbar \e \dot \rho =\lin_s(\rho),\quad (\e>0)
    \ee
where the slowly varying parameter $s=\e t$, having the physical dimension of
an energy, is viewed as the slow
clock. $\lin_s$ is the changing Lindblad operator
    \be\label{eq:lindblad}
    \lin(\rho)= -i[H,\rho]-\hbar\gamma (P_-\rho P_++P_+\rho P_-);
    \ee
$H$ is the Hamiltonian, which for a generic near crossing is
given in Eq.~(\ref{eq:H}); $P_\pm=\ket{\pm}\bra{\pm}$ are the two spectral
projections of $H$; finally,
$\gamma
>0$ is the dephasing rate \cite{t2}. $\gamma=0$ is the case considered by
Landau and Zener. In both cases transitions between the ground and the excited states only occur
because the generator of the dynamics depends on $s$.
The tunneling probability
\be
T
 = \tr(\rho P_+)(\infty),\quad \big(\,\rho(-\infty) =
P_-(-\infty)\,\big) \ee is the error in fidelity of the ground
state.

The adiabatic tunneling formula with dephasing, which we shall
derive below, is \cite{error}
    \be\label{eq:DLZ}
   T
   =\frac{\e\hbar}{2 g_0^2}\, Q\Bigl(\frac {\hbar\gamma}{g_0} \Bigr)+O(\e^2),
   \ee
where $Q$ is the algebraic function (shown in the figure)
   \be\label{eq:Q}
   Q(x)=
   \frac{\pi}{2}\frac{x(2+\sqrt{1+x^2})}{\sqrt{1+x^2}(\sqrt{1+x^2}+1)^2}.
    \ee

\begin{figure}[htb]
\centering
\includegraphics[width=0.6\textwidth]{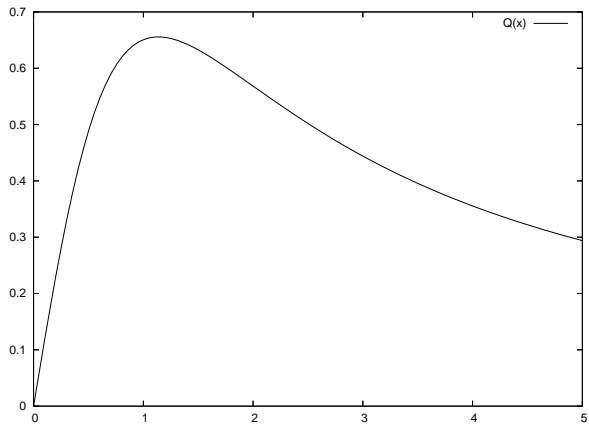}
\caption{The function $Q(x)$. The argument is the ratio of
dephasing rate to the minimal gap. The function has a maximum at
$x=1.13693$} \label{fig:q}
\end{figure}

Few remarks about this result are in order:
    \begin{itemize}
     \item
    The adiabatic limit means that $\sqrt{\hbar\e}$ is the smallest
energy scale in the problem and in particular, $
    \e \ll \hbar\gamma^2 $.
When this fails, the error terms in Eq.~(\ref{eq:DLZ}) need not be
small compared to the leading term.
        \item
        When the dephasing is weak, $\hbar\gamma\ll g_0$, Eq.~(\ref{eq:DLZ}) reduces to
    \be\label{eq:WD}
   T\approx
    \frac{3 \pi}{16}\cdot\frac{\e\gamma\hbar^2}{g_0^3}.
    \ee
This term has the same form as one of the tunneling terms found by
Shimshoni and Stern \cite{shimshoni} in a (different) model where
the two-level system is dephased by noise. The method they use
cannot give the overall constant $3\pi/16$ \cite{Berry},  nor does
it allow investigating the full range of $\hbar \gamma/g_0$.
 \item 
When $\hbar\gamma\gg g_0$ Eq.~(\ref{eq:DLZ}) reduces to
    \be\label{eq:gG}
    T\approx     \frac{\pi\e}{4 g_0\gamma}.
    \ee
This may be understood as a manifestation of the quantum Zeno
effect \cite{zeno}: The dephasing term in the Lindblad generator
can be interpreted as monitoring the state of the system at rate
$\gamma$. This suppresses transitions between the states.
    \end{itemize}

Eq.~(\ref{eq:DLZ}) follows from, and is  a special case of, a more
general and basic formula for the tunneling when the adiabatic
evolution takes place on a finite interval of (slow) time
$[s_0,s_1]$ and one also allows $\gamma(s)$ to be time-dependent
    \be\label{eq:res}
    T = 2 \e\hbar^2 \int_{s_0}^{s_1}  \gamma(s) \frac{ \tr(P_+
    {\dot{P}_-}^{\,2} P_+)}{g^2(s) +\hbar^2\gamma^2(s)}\, ds+O(\e^2 ),
    \ee
Here $g(s)$ is the instantaneous gap in $H(s)$,
    \be
    g^2(s)=s^2+g_0^2.
    \ee
The positivity of the integrand in Eq.~(\ref{eq:res}) when $\gamma>0$
makes the tunneling irreversible. This changes the characteristics
of the $\e$ dependence of $T$ from exponentially small in
Eq.~(\ref{eq:LZT}) to linear in Eq.~(\ref{eq:DLZ}).  The
Landau-Zener formula, Eq.~(\ref{eq:LZT}), is buried in the error
terms of Eq.~(\ref{eq:res}).

Eq.~(\ref{eq:res}) reduces the tunneling problem to integration.
In the case where $\gamma$ is constant and $s$ runs from $-\infty$
to $\infty$ the numerator Eq.~(\ref{eq:res}) is simply
    \be
    \tr(P_+
    {\dot{P}_-}^{\,2} P_+)
    =\frac{g_0^2}{4g^4(s)}.
    \ee
Elementary algebra then leads to Eq.~(\ref{eq:DLZ}) with
   \be
   Q(x)= x \int_{-\infty}^\infty (t^2+1)^{-2} (t^2+1+x^2)^{-1}dt.
    \ee
The integral can be evaluated explicitly to give Eq.~(\ref{eq:Q}).

The key idea behind the derivation of the adiabatic
tunneling formula, Eq.~(\ref{eq:res}), is a geometric view of the
spectral projection as {\em adiabatic invariants}. The evolution
of {\em observables} is governed by the adjoint of the Lindblad
generator, $\lin^*$, (this is the Heisenberg picture), where the adjoint 
refers to the Hilbert-Schmidt inner product. In
particular, the adjoint of the dephasing Lindblad operator of
Eq.~(\ref{eq:lindblad}) acting on the observable $A$ is given by
(from now on we set $\hbar=1$)
%%%%%
    \be
    \lin^*(A)= i[H,A]-\gamma (P_-A P_++P_+A P_-),
    \ee
It differs from Eq.~(\ref{eq:lindblad}) by the replacement of $i$
by $-i$. As we shall now see an instantaneously stationary
observable $A(s)\in \Ker(\lin^*_s)$ that has no motion in
$\Ker(\lin_s)$ is an {\em adiabatic invariant}.
More precisely,
%%%%%%
\begin{thm}
Let $A(s)$ be an observable which lies in the instantaneous kernel
of $\lin^*_s$, i.e.
    \be\label{eq:kernel}
    \lin^*_s\big(A(s)\big)=0
    \ee
and suppose that, in addition, the linear equation
    \be\label{eq:X}
    \dot A(s)=\lin^*_s(X(s))
    \ee
admits a solution $X(s)$. Then one has
    \be\label{ai}
    \tr(A(s)\rho_\e(s))\big|^{s_1}_{s_0}=
    \e\tr(X(s)\rho_\e(s))\big|^{s_1}_{s_0}-
    \e\int^{s_1}_{s_0}\tr(\dot X(s)\rho_\e(s))\,ds,
    \ee
where $\rho_\e(s)$ is a solution of the adiabatic Lindblad
evolution. $A(s)$ is an adiabatic invariant in the sense that its
expectation is conserved up to a small error, $O(\e )$, given by
the right hand side of Eq.~(\ref{ai}) whereas the change in a generic observable is $O(\e^{-1})$ and in the Lindblad generator is $O(1)$.
\end{thm}

The identity, Eq.~(\ref{ai}), readily follows from
\begin{align}
\frac{d}{ds}\tr(A(s)\rho_\e(s))
&=\tr(\dot A(s)\rho_\e(s))+\tr(A(s)\dot\rho_\e(s))\nonumber\\
&=\tr(\lin_s^*(X(s))\rho_\e(s))+\e^{-1}\tr(A(s)\lin_s(\rho_\e(s)))\nonumber\\
&=\tr(X(s)\lin_s(\rho_\e(s)))+\e^{-1}\tr(\lin_s^*(A(s))\rho_\e(s))\nonumber\\
&=\e\tr(X(s)\dot \rho_\e(s))
\end{align}
and integration by parts.

Eq.~(\ref{eq:X}) may be interpreted as a condition that $A(s)$ undergoes parallel transport: The equation has a solution provided $\dot A(s)\in \Range\, (\lin_s^*)$ which is the case if $A(s)$ has no motion in $\Ker\, (\lin_s)$.

It is straightforward to verify that the instantaneous spectral
projections $P_j(s)$ of a dephasing Lindblad generator are
adiabatic invariants in the sense of the theorem. Evidently, 
$\lin^*_s\big(P_+(s)\big)=0$.
Moreover, Eq.~(\ref{eq:X}) is solved by
    \be\label{eq:XX}
    X(s) = -i\sum_{k\neq j} \frac{P_k \dot{P}_+ P_j}{e_k - e_j +i\gamma}
    \ee
with $e_\pm$ the two eigenvalues of $H$. To see this note first
that $X(s)$ is  purely off-diagonal \cite{remark} by construction
and so is $\dot P_+$, namely
    \be\label{pid}
    \dot P_+=P_-\dot P_+ P_+ +P_+ \dot P_+P_-.
    \ee
This follows form $P_+=P_+^2$, which implies $\dot P_+=\dot P_+
P_+ + P_+ \dot P_+$ and in turn  $P_\pm \dot P_+P_\pm=0$.
The equality of the off-diagonal components of Eq.~(\ref{eq:X}) follows from
    \be
    \lin^*(P_k A P_j) = i(e_k - e_j +i \gamma)P_k
A P_j, \quad (k,j =\pm, k\neq j)\,.
    \ee

The probability of leaking out of the instantaneous ground state
is given by Eq.~(\ref{ai}) with $A(s)=P_+(s)$. Eq.~(\ref{eq:res})
then follows by appealing to the adiabatic theorem
\cite{joye} which allows to replace the instantaneous state
by the instantaneous projection on the right hand side of
Eq.~(\ref{ai}),
    \be
    \rho_\e(s)= P_-(s)+O(\e),
\label{adiabat}
    \ee
uniformly in $s_0$, $s$. The rest is simple algebra. 

For the convenience of the reader we include a proof of Eq.~(\ref{adiabat}). 
Let $U_\e(s,s_0)$ be the propagator for the differential equation 
(\ref{eq:L}), whence $\rho(s)= U_\e(s,s_0)P_-(s_0)$. We recall that the 
solution of its inhomogeneous variant, $\dot x=\e^{-1}\lin_s(x)+y$, is given 
by the Duhamel formula
\be\label{eq:duh}
x(s)=U_\e(s,s_0)x(s_0)+\int_{s_0}^{s} U_\e(s,s')y(s')ds'.
\ee
The remainder to be estimated, $r(s)=\rho(s)-P_-(s)$,
satisfies $\dot r=\e^{-1}\lin_s(r)-\dot P_-$, because of $\lin_s(P_-)=0$. 
Before applying Eq.~(\ref{eq:duh}), we observe that the equation 
$\dot P_-(s)=\lin_s(X(s))$
admits Eq.~(\ref{eq:XX}) as a solution upon replacing  $i$, $\dot P_+$ by 
$-i$, $\dot P_-$. 
The differential equation thus becomes
\[
(r-\e X)^{\cdot}=\e^{-1}\lin_s(r-\e X)-\e \dot X,
\]
resulting in 
\[
r(s)-\e X(s)=
U_\e(s,s_0)(r(s_0)-\e X(s_0))-\e\int_{s_0}^{s}U_\e(s,s')\dot X(s')ds'.
\]
Since $\lin_s$ is dissipative, i.e. $\tr(\rho \lin_s(\rho))\le 0$, we obtain
$\|U_\e(s,s_0)\|\le 1$, ($s\ge s_0$). Together with $r(s_0)=0$ we conclude that
$r(s)=O(\e)$, as claimed. The uniformity follows from decay: $X(s)=O(s^{-3})$, 
$\dot X(s)=O(s^{-4})$, ($s\to\pm\infty$).   

In conclusion: We have introduced a class of adiabatically
changing dephasing Lindblad operators which allowed us to
calculate the tunneling  in a generic two-level crossing and
extend the Landau-Zener tunneling to dephasing Lindbladians with
arbitrary dephasing rate. Dephasing makes the tunneling
irreversible and so fundamentally different from tunneling in the
unitary setting. This irreversibility is responsible for the
difference in the functional form of the tunneling formulas.

{\bf Acknowledgment.} This work is supported by the ISF and the
fund for Promotion of research at the Technion. The last two authors are
grateful for hospitality at the Physics Department at the Technion, where
most of this work was done.
Useful discussions with A.~Keren and E.~Shimshoni are acknowledged.


\begin{thebibliography}{99}
\bibitem{history} Landau, who used semiclassical methods
did not actually  attempt to calculate the multiplicative overall
factor in front of the exponential in Eq.~(\ref{eq:LZT}).
Fortunately, this factor happens to be unity. Zener solved the
differential equation exactly in terms of Weber functions and
derived Eq.~(\ref{eq:LZT}) exactly. Zener was aware of Landau's
solution but for some reason, incorrectly, believed that Landau
missed a factor of $2\pi$ in the exponent.
\bibitem{t2} $\gamma^{-1}$ is the dephasing time commonly denoted by  $T_2$.
\bibitem{error} Since the problem has two
dimensionless parameters: $\hbar\e/(\hbar^2\gamma^2+ g_0^2)$ and
$\hbar\gamma/g_0$ it is not obvious what is the correct
dimensionless expression corresponding to $O(\e^2)$.  The correct
interpretation of $O(\e)$ turns out to be the product $
O\left(\frac{\hbar\e}{\hbar^2\gamma^2+g_0^2}\,\frac {\hbar \gamma}
{g_0}\right)$ which can be verified by Eq.~(\ref{eq:res}).
\end{thebibliography}
\end{document}